# BLUETOOTH NAVIGATION SYSTEM USING WI-FI ACCESS POINTS

Rohit Agrawal[1] and Ashesh Vasalya[2]

[1]School of Electronics Engineering, VIT University, Vellore
`rohit4849@gmail.com`
[2]School of Electrical Engineering, VIT University, Vellore
`asheshvasalya@yahoo.in`

*ABSTRACT*

*There have been various navigation and tracking systems being developed with the help of technologies like GPS, GSM, Bluetooth, IR, Wi-Fi and Radar. Outdoor positioning systems have been deployed quite successfully using GPS but positioning systems for indoor environments still do not have widespread deployment due to various reasons. Most of these use only a single technology for positioning but using more than one in cooperation with each other is always advantageous for obtaining greater accuracy. Particularly, the ones which use Bluetooth are better since they would enhance the scalability of such a system because of the fact that this technology is in use by the common people so it would always be easy to track them. Moreover it would also reduce the hardware installation cost to some extent. The system that has been introduced here uses Bluetooth primarily for positioning and tracking in combination with Wi-Fi access points. The reason that makes the commercial application of such a system easier and cheaper is that most of the localized areas today like college campus, offices are being provided with internet connectivity using these access points.*

## 1. INTRODUCTION

With the large scale expansion of campuses or premises of colleges, hospitals, corporate offices, hotels and all sorts of commercial and non-commercial buildings, precise positioning has gained a lot of importance not only to save time but also to get rapid access to everybody everywhere. There are a lot of organizations which have some restricted areas which they would like to make accessible to only a few people, this facility can also be provided by such tracking systems. Protecting secured networks from intrusion or maintaining a record of all the movements of the network assets are some other advantages of these systems.

Although such tracking systems have such varied applications but the most commercially viable ones in the present scenario are those which make use of the existing wireless infrastructure so as to reduce the equipment and installation cost considerably. Wi-Fi tracking seems to be a plausible solution and significant work has been done in this regard but there are a lot of drawbacks associated with them.

The system introduced in this paper is based on two most popular wireless technologies namely Bluetooth and Wi-Fi. The sole motive of using Bluetooth is that it is widely available in the mobile phones now days and is designed for low power consumption compared to Wi-Fi. Moreover, we make use of Bluetooth – Wi-Fi Gateways which are integrated nodes and can communicate with both Bluetooth and Wi-Fi enabled devices and finally, we have Wi-Fi access points which are already a part of the organization's infrastructure to provide internet connectivity.

The two concerned wireless technologies are being discussed in brief next.

## 1.1 Bluetooth

Bluetooth is an open wireless technology initially considered for replacement to the RS232 serial cables, now finds varied applications in consumer as well as industrial devices and practises [13]. It is based on frequency-hopping spread spectrum and uses a master-slave structure for establishing connection. One master can connect to seven slaves at a time forming a piconet network.

The functionality of a Bluetooth device is defined on the basis of the Bluetooth profiles, it has implemented. There are lot of such profiles defined, some of them are Advanced Audio Distribution Profile (A2DP) for streaming stereo audio quality from a media source to a sink, Human Interface Device Profile (HID) which defines the protocols, procedures and features to be used by Bluetooth HID such as keyboards, pointing devices, gaming devices and remote monitoring devices, Video Distribution Profile (VDP) which defines how a Bluetooth enabled device streams video over Bluetooth wireless technology etc .

## 1.2 Wi-Fi

Wi-Fi is wireless technology based on the IEEE 802.11 standards [14] and is synonymous with Wireless Local Area Network (WLAN) since it is seen as a replacement to the cabling required in early Local Area Networks.

Its major application is to provide high speed internet connectivity to the Wi-Fi enabled devices while being in range of wireless network connected to the internet. This kind of network is implemented with the help of access points (hot spots) and routers. Wi-Fi also uses the same radio frequencies as that of Bluetooth but requires more power for operation and thereby provides higher bit rates and better range of communication.

The rest of the paper is divided into various sections. Section 2 describes the related work done before which is important to understand the relevance of the proposed solution. System components have then been explained in Section 3 prior to the description of the system implementation in Section 4.

## 2. RELATED WORK

The most widely used technology for this application is definitely Global Positioning System(GPS) which is a world-wide satellite based communication system [1]. However, GPS cannot be used indoors because a GPS receiver usually fails if line of sight visibility to the satellites is lost. Assisted GPS may work indoors but it has its own limitations with respect to network assistance. Being global in nature, GPS gives the latitude and longitude details of a GPS receiver but the kind of tracking that we are talking about is in a localized area comprising of both indoor and outdoor locations. Systems relying on cellular networks [2] have to be owned and administered by the service provider and not the organization and may need subscription fee as well.

The systems that completely depend on Wi-Fi consider either laptops or mobile phones as their mobile nodes to track down the owners of these devices. However, using laptops for this purpose may give incorrect results since a person is not carrying his laptop everywhere he goes and they might not even be switched ON every time. At the same time, using mobile phones for tracking also does not seem plausible since Wi-Fi function is available only in expensive high end smart phones. A mere connection of a person with an access point can help us to locate him by finding out the location of the access point on the basis of its MAC ID but this would be in a range of around 30 meters (the range of a normal Wi-Fi access point). So, most of these systems are based on Received Signal Strength Indicator (RSSI) [2] [10] and Signature Matching [3] or Fingerprinting [4]. The former are based on propagation models that relate the signal strength directly to the distance from the Wi-Fi access point (AP) using, for example the Friss formula [5] and can work for stagnant or slow moving assets. When the mobile devices are moving fast, the signals tend to change rapidly and sometimes, obstacles will also affect the signal values making the distance prediction either difficult or inaccurate. In the latter ones, a database of different signal strengths for different locations (called signatures) is created in the server and the person is positioned by comparing them with the current signal strength. Any change in the arrangement of obstacles calls for retraining [3]. There might even be different signal strength values for different types of devices (laptop or mobile) kept at same distance from access point. So these systems require bulky databases, complex processing or training in the servers to determine the correct position. Above were some disadvantages of the systems relying only on Wi-Fi for tracking which make them inefficient. There were some systems developed based on Bluetooth as well. One such system was established at the Aalborg Zoo where Bluetooth access points were installed throughout [6]. At the entrance, parents rent BlueTags BodyTag for their children and register the required information along with contact details. The parents make enquiry about their children through an SMS code and get response with the help of the tracking software. One major drawback of such a system was that apart from the installation cost of the access points, the cost of the bluetags also was to be borne. This also put a constraint that only a limited no. of children could be tracked if the no. tags were limited. However in our system, the Bluetooth enabled mobile phones act as the Bluetooth nodes. Only the installation of a minimum no. of Bluetooth access points will be needed which will work in conjunction with the Wi-Fi access points that are already existent today in the campuses to provide internet connectivity. The system illustrated in [15] has enhanced BT positioning accuracy upto 1.5 meters but only with modifications in the core stack which may not be possible with the BT enabled mobile phones of users.

There were a few systems depending on two technologies simultaneously on Wi-Fi and Zigbee [7] but Zigbee is not a common consumer technology. In [8], fusion based system on Wi-Fi and Bluetooth is explained which performs better positioning than system relying only on Wi-Fi but here there did not exist an internetwork between the two, it tried to fuse separately obtained results.

## 3. SYSTEM COMPONENTS

### 3.1. Bluetooth enabled mobile phone

All the persons in the network own a Bluetooth enabled mobile phone through which they connect to the network, which they use to make a request to track others and also for themselves to be located by others.

### 3.2. Bluetooth (BT) Access Points

A BT Class 3 device with a range of around 10 meters may also be called as reader node is a Bluetooth transceiver present for an immediate connection to the mobile phone. These access points are needed to incorporate the core Bluetooth protocols (mandatorily present in all BT devices) like Link Management Protocol (LMP), Logical Link Control and Adaptation Protocol (L2CAP) and Service Discovery Protocol (SDP). Additionally, there is need of Radio Frequency Communications (RFCOMM) protocol which provides for binary data transport by creating a virtual serial data link.

### 3.3. Gateway Nodes

These are integrated Bluetooth and Wi-Fi nodes which act as a gateway to connect the bluetooth network to the Wi-Fi network. Both these wireless technologies operate in the 2.4 GHz radio frequency range allowing for the option of a shared antenna system which has been successfully incorporated in BCM4325 chip featuring Broadcom's InConcert technology, which consists of sophisticated software algorithms and hardware mechanisms that enable collaborative co-existence between Wi-Fi and Bluetooth and thereby paving way for entirely new network applications.

### 3.4. Wi-Fi Access Points

They allow wireless devices to connect to a wired network like that of an organization's Local Area Network to provide access to the shared information and also internet connectivity. In our system, they acquire tracking information from other devices and sent it to the server and vice versa.

### 3.5. Server

The Server in our system apart from performing various network functions also stores information of the network's assets i.e. their bluetooth addresses or device names. Moreover it also contains a list of different BT APs with their address and corresponding physical location in the form of a look up table shown in Table 1.

## 4. SYSTEM IMPLEMENTATION

With the development of wide variety of applications for the operating systems of mobile devices and their easy availability for free, the applications of mobile phones has also increased manifold. Several applications [9] [11] (mostly based on Java platform) have also been developed for the purpose of tracking. As soon as a person belonging to that organisation enters inside the premises, he can connect to the BT network via reader nodes (BT Access Points). When he wants to know the position of another person, he makes use of such a tracking application provided in all the mobile phones. He either sends the name of the person or the bluetooth device address (48 bit unique BT id) to the nearby reader node via bluetooth. This information hops between different reader nodes to finally reach the Wi-Fi AP through the Gateway node. Fig. 1 illustrates the network of these access points.
Snippets of structure definitions of the code have been mentioned with their description.

```
Struct mobinfo
{
Long bsid;
String name;
String another_info;
}
```

The above structure contains the information of the person to be tracked wherein the Bluetooth id is enough to identify the person uniquely but supporting information like name and another_info are sent to be sure. The reader node captures this information from the mobile device, its software adds the originator's static id (to get back to the originator with the tracking information) and sends it to the nearby reader node with the help of RFCOMM protocol. The revised structure, after the data has been modified by the reader node looks like this:

```
Struct modified_mobinfo
{
Long bsid;
String name;
String another_info;
Static int originator_id;
}
```

This updated information travels node by node via bluetooth to reach the gateway node which is present near the Wi-Fi access point. Since this node is integrated with both Wi-Fi and Bluetooth, it communicates the information to the Wi-Fi AP and is then transferred to the organization's local area network to the server through an ethernet connection. Fig. 2 shows the complete system layout to illustrate the transfer of information between different network components.

Apart from this tracking information, the server also handles a lot of other data as well so we need packet extractor application to extract the packet related to the tracking information at the server end for further processing. The server now broadcasts this information to all the access points then to all the reader nodes to locate the person's Bluetooth device. All the reader nodes now try to find this device by making an attempt to establish a connection with the known BT ID or the device name. As soon as a connection is made to the required person, the particular reader node that made the connection sends its ID to the server. The server refers to the look up table to obtain the physical location of the access point and gives this information to the originator who made the tracking request and also update the information in case the tracked person is in motion by identifying current connection with the reader node. In this way, positioning can be achieved within a radius of 10 meters.

### 4.1. Figures and Tables

Table 1. Lookup Table for BT Access Points.

| S.No. | Bluetooth AP ID | Physical Location |
|---|---|---|
| 1. | 00:0C:25:14:67:1E | C Block-Indoor Court |
| 2. | 00:4A:12:81:1A:BD | Digital Library |
| 3. | 00:82:44:A6:BB:10 | Amphitheatre |
| 4. | 00:86:31:EA:89:22 | Canteen |
| 5. | 00:75:C2:78:56:E1 | Student Council Center |

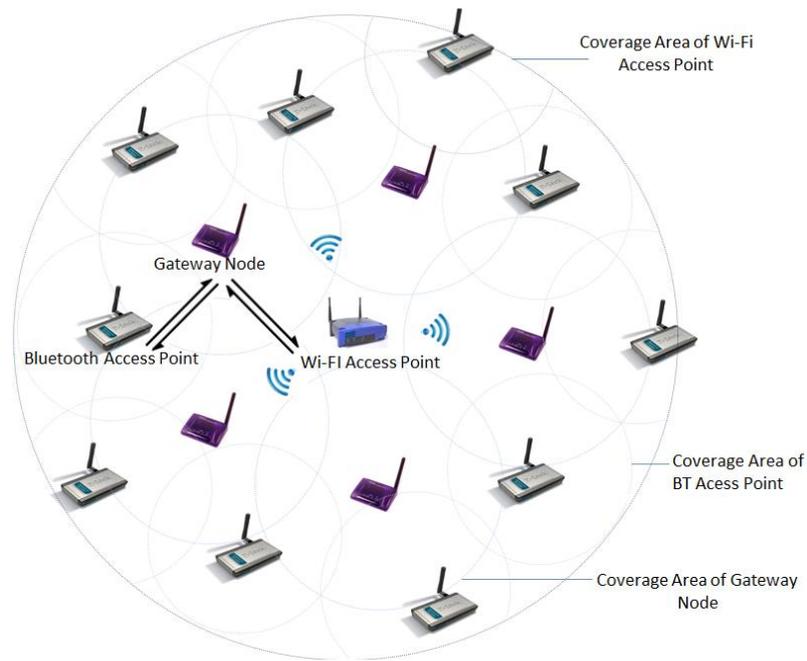

Figure 1. Network of Different Access Points

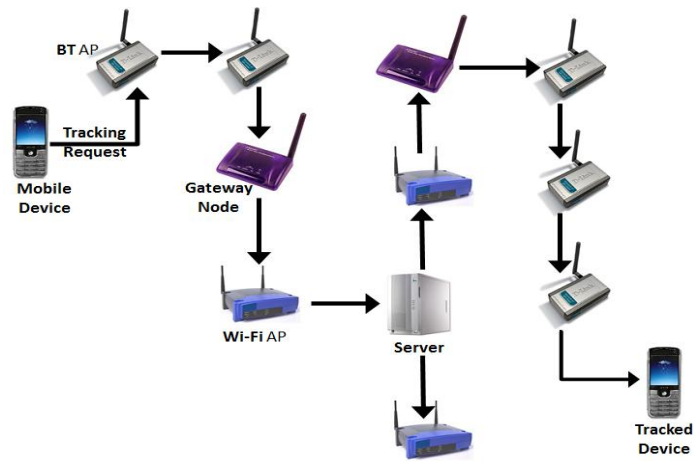

Figure 2. System Layout

## 5. CONCLUSION

The paper mainly introduces a new tracking system that relies on Bluetooth and Wi-Fi simultaneously. The main aim is not only to make the best utilization of the existing infrastructure available in an organization but make the deployment of the system most commercially viable by using technologies that are already available with the consumers. In the proposed system, the authors have achieved a tracking accuracy of 10 meters which is the range of the Class 3 BT devices and is the most prevalent class of BT devices.

Another advantage of using such a system lies in its use of mobile phones for tracking. With development of mobile phone apps on almost all platforms at such a large scale, there is a lot of scope of adding new capabilities and features to this system. One such example would be to generate a map/path based on the location of the user making request and the location tracked which would further ease the process of positioning.

## Authors

Rohit Agrawal is currently working as a Graduate Engineer Trainee with Magneti Marelli Powertrain India Ltd. in the Manufacturing Department. He is involved in various automation projects in the Process and Testing areas. He completed his engineering graduation from VIT University, Vellore in the year 2011 in Electronics and Communication Department.

He did an internship project on Power Optimization in Numeric LCD's at India Institute of Technology, Bombay. He carried out his final year project in Delphi Automotive – Technical Center India on Automotive Diagnostics. He is the author of two papers Springer and IEEE respectively illustrating his above project works. He is also involved in an incubated project under Technopreneurship Promotion Programme (TePP) by Department of Science and Technology. His areas of interest include Embedded Systems Design, In-vehicle Networking and Wireless Communication.

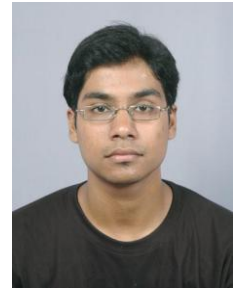

Ashesh Vasalya is currently pursuing final year of B.Tech degree in Electronics and Instrumentation Engineering from Vellore Institute of Technology University (VIT-U), Vellore, India, 2008-12.

He has worked for Mini Baja SAE India 2011 and has completed his academic internship on the Implementation of MOST25 Protocol at TIFAC-CORE in automotive Infotronics, Department of Science and Technology, Govt. of India. He is currently the board member of ISOI (Instrument Society of India) VIT-U branch and the member of SAE (Society of Automotive Engineers). He is currently doing his major project on Self Optimization of 3G and 4G LTE Systems. He is also the IEEE author of a paper on Robotics. His areas of interests are Advanced Robotics, Advanced Embedded Device Driver and Automotive Communication Protocols.

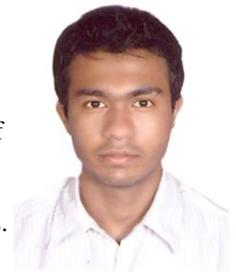